# The qubit information logic theory for understanding multi-qubit entanglement and designing exotic entangled states

Zixuan Hu and Sabre Kais*

*Department of Chemistry, Department of Physics, and Purdue Quantum Science and Engineering Institute, Purdue University, West Lafayette, IN 47907, United States*
*Email:* kais@purdue.edu

Abstract: We develop a "qubit information logic" (QIL) theory that uses the "qubit information equation" (QIE) and logic to describe the correlation behaviors of multi-qubit entanglement. Introducing the "global information status" and "local information availability", the QIL gives an alternative and natural interpretation of the "spooky action" and the quantum no-communication theorem. Compared to the conventional entropy-based entanglement theories, the QIL directly describes the correlation of each possible pair of qubits and how the correlation changes when other qubits are measured. This makes the QIL more advantageous in describing the correlation properties of multi-qubit entanglement, which is illustrated by studying the dormant entanglement phenomenon. The QIL theory's usefulness is further demonstrated by designing an exotic quantum state where two qubits can be entangled but not correlated in any arbitrary basis. Overall the QIL provides an alternative and intuitive understanding of multi-qubit entanglement that is, compared to the conventional theories, directly focused on the correlation behaviors between qubits and thus more suitable for designing exotic quantum states that may be used in quantum algorithms.

## 1. Introduction

Quantum computation and quantum information have received enormous attention and advanced rapidly in both theoretical and technological directions. Over the last thirty years, many sophisticated theories and state-of-the-art technologies have been developed [1-32]. Entanglement being a unique quantum phenomenon has been widely studied and used as a valuable resource in quantum information science. In particular, model quantum systems such as the Bell states and the GHZ (Greenberger–Horne–Zeilinger) state exhibit quantum-only correlation behaviors between qubits when they are measured in different bases [33-35]. Exploiting the non-locality of entanglement on such quantum systems, novel quantum communication protocols including quantum teleportation [36, 37] and quantum key distribution [38-43] have been designed and realized. The conventional theories of entanglement are mostly entropy-based [44-46], which give a mathematical measure of entanglement that can be then used to describe the complexity of quantum circuits [47] used in quantum computation and quantum information. However, from a practical perspective of developing quantum algorithms with certain functionalities, the measurement statistics and correlation behaviors of qubits in e.g. ansatzes and output states directly determine the performance and final results of quantum computing tasks – yet these properties are not well described by the conventional theories [25]. For example, 1. two qubits can be entangled but have no correlation when measured in any basis (see Section 2.5 below); 2. describing the correlation behaviors of all possible pairs of qubits in a multi-qubit state is difficult with the conventional theories; 3. entanglement can depend on the partitioning of the space [25] and the basis choice of external systems [48].

To address these limitations, in this work we develop a new "qubit information logic" (QIL) theory that uses the "qubit information equation" (QIE) and logic to directly describe correlations between qubits in different bases. Firstly, the QIL theory is completely consistent with the conventional understanding of entanglement as it provides a correct and intuitive interpretation of the spooky action and the quantum no-communication theorem. Secondly, the QIL theory shows its advantages in describing the correlation behaviors of multi-qubit entanglement as illustrated by the dormant entanglement states. In particular, the QIE together with logical reasoning on the qubit values completely describes how each possible pair of qubits should correlate when measured in the current and the Hadamard-rotated bases, and how this correlation behavior changes when other qubits are measured. Utilizing the advantages of the QIL theory, we proceed to resolve an apparent paradox of the dormant entanglement phenomenon. Finally we demonstrate the QIL theory's usefulness by designing an exotic quantum state where two qubits are entangled but not correlated in any arbitrary basis.

## 2. The qubit information logic (QIL) theory

**2.1 The qubit information equation (QIE).** We start with the simplest entanglement, the Bell state:

$$|\varphi_1\rangle = \frac{1}{\sqrt{2}}\left(|00\rangle_{12} + |11\rangle_{12}\right) = \frac{1}{\sqrt{2}}\left(|++\rangle_{12} + |--\rangle_{12}\right) \quad (1)$$

where $|\pm\rangle = \frac{1}{\sqrt{2}}(|0\rangle \pm |1\rangle)$; and the subscripts outside the bracket, e.g. $|00\rangle_{12}$, represent the numbers used to identify the qubits. What distinguishes $|\varphi_1\rangle$ from a classical system is that there is perfect correlation between the two qubits $q_1$ and $q_2$ in both the current $\{|0\rangle, |1\rangle\}$ basis and the Hadamard-rotated $\{|+\rangle, |-\rangle\}$ basis. The mathematical form and the correlation behavior of $|\varphi_1\rangle$ are very simple. However, here to introduce the new qubit information logic (QIL) theory we pretend to encounter $|\varphi_1\rangle$ for the first time and ask the question: why do the two qubits in $|\varphi_1\rangle$ have perfect correlation in both the $\{|0\rangle, |1\rangle\}$ and the $\{|+\rangle, |-\rangle\}$ bases? To answer this question we propose the qubit information equation (QIE) of $|\varphi_1\rangle$ is:

$$q_1(c) \oplus q_2(c) = 0 \quad (2)$$

where the values of the two qubits in the current basis ("$c$" means "current") are related by an equation; $\oplus$ means addition modulo 2. If we first measure $q_1$ in the current basis to get $q_1(c)$'s value, then regardless if we get $q_1(c) = 0$ or 1, by Eq. (2) we know that $q_2(c)$ must have the same value as $q_1(c)$, so measuring $q_2$ in the current basis must give a deterministic value. Similarly, if we determine $q_2(c)$ first, then $q_1(c)$ has a deterministic value. Consequently Eq. (2) describes the perfect correlation of the two qubits if measured in the current basis. What may not be obvious is that Eq. (2) also describes the perfect correlation if the qubits are measured in the Hadamard-rotated basis. Suppose we rotate $q_1$ by a Hadamard gate and then measure it, then its value in the Hadamard-rotated basis $q_1(h)$ is determined ("$h$" means "Hadamard"), then due to complementarity, $q_1$'s information in the current basis, i.e. $q_1(c)$'s value, is lost (fundamentally random). Now because by Eq. (2), $q_2(c)$'s value is equal to $q_1(c)$'s value, the loss of $q_1(c)$ means that $q_2(c)$'s value is lost too – again by complementarity this then means $q_2(h)$ is now deterministic if we now choose to measure it. So the QIE in the current basis as in Eq. (2) effectively implies another QIE in the Hadamard-rotated basis:

$$q_1(h) \oplus q_2(h) = \text{"deterministic value"} \tag{3}$$

and thus the two qubits must be also perfectly correlated in the Hadamard-rotated basis. The exact value of $q_1(h) \oplus q_2(h)$ is 0 if we started with $|\varphi_1\rangle = \frac{1}{\sqrt{2}}(|00\rangle_{12} + |11\rangle_{12})$; is 1 if we started with $|\varphi_2\rangle = \frac{1}{\sqrt{2}}(|00\rangle_{12} - |11\rangle_{12})$: note that $|\varphi_1\rangle$ and $|\varphi_2\rangle$ share the same QIE in the current basis as in Eq. (2) but have different deterministic values for the QIE in the Hadamard-rotated basis as in Eq. (3). If we combine the QIEs in both "$c$" and "$h$", the complete QIE that uniquely identifies $|\varphi_1\rangle$ is:

$$\begin{cases} q_1(c) \oplus q_2(c) = 0 \\ q_1(h) \oplus q_2(h) = 0 \end{cases} \tag{4}$$

**2.2 Global information status and local information availability.** Here we propose that the QIEs in Eqs. (2), (3) and (4), and the QIL reasoning allow a natural interpretation of the "spooky action" and the quantum no-communication theorem [36]. Consider the Bell state $|\varphi_1\rangle = \frac{1}{\sqrt{2}}(|00\rangle_{12} + |11\rangle_{12})$, its QIE in Eq. (4) describes the correlation of $q_1$ and $q_2$'s information in both the current basis and the Hadamard-rotated basis. In particular, $q_1(c) \oplus q_2(c) = 0$ means that measuring $q_1$ in the current basis determines $q_2$'s value in the same basis (and vice versa); while $q_1(h) \oplus q_2(h) = 0$ means that measuring $q_1$ in the Hadamard-rotated basis determines $q_2$'s value in the same basis (and vice versa). Consequently, measuring $q_1$ or $q_2$ in a given basis has an instantaneous effect on the other qubit's measurement statistics – this is the "spooky action". Now to explain the spooky action with the QIL theory we propose that each qubit has its own

"information status" that specifies whether the qubit's information has been determined or not, and if determined then in what basis. Importantly, this information status is known by Nature as a global property not restricted by the speed of light. Before any measurement is made on $|\varphi_1\rangle$, each qubit's information status is "undetermined", and the QIE in Eq. (4) specifies the conditions that the qubits must satisfy once their information statuses become determined in the respective basis. After a measurement in the current basis has been made on $q_1$, its information status changes to "$q_1(c)$ is determined", which then changes $q_2$'s information status to "$q_2(c)$ is determined", because we have to satisfy the QIE $q_1(c) \oplus q_2(c) = 0$. Now because the information status is known by Nature as a global property, $q_2$'s information status is instantaneously updated by $q_1$'s measurement, regardless how far away $q_2$ is. Similar situations happen if we measure $q_2$ first, or measure in the Hadamard-rotated basis: the information status of the other qubit is always instantaneously updated by the measurement. This is the QIL interpretation of the spooky action.

In contrast to the information status being a global property, the basis in which the qubit is measured and the value as measured, i.e. the information itself, is subject to the "information availability" that is a local property restricted by the speed of light. The QIL thus explains the quantum no-communication theorem as follows: the information status change of one qubit instantaneously changes the other qubit's information status (spooky action), but being physical observers located near $q_2$ trying to detect such changes, we need the information itself on what actual change has happened on $q_1$, which then involves information availability. Then because information availability is local, we need to wait for the communication from $q_1$, which is restricted by the speed of light, therefore we have the quantum no-communication theorem.

There are two core ideas in the QIL interpretation of the spooky action and the quantum no-communication theorem. Firstly, the information status of each qubit in $|\varphi_1\rangle$ is "undetermined" because it reserves the possibility of being measured and "determined" in a chosen basis. Secondly, the information status of each qubit in $|\varphi_1\rangle$ has two complementary information forms ($q(c)$ and $q(h)$) that can potentially be determined: if one form is determined, the other one is "lost" (fundamentally random). Due to the first idea of information status being undetermined, we can have the instantaneous change of information status that leads to the spooky action. Due to the second idea of having two complementary information forms, being near $q_2$ we cannot know what actual change has been made by the measurement on $q_1$ without communication, which leads to the quantum no-communication theorem. Both ideas are fundamentally quantum as classical systems cannot have undetermined information or complementary information forms. In a way, the QIL interpretation of the spooky action and quantum no-communication theorem is essentially a criterion of entanglement based on information: entanglement exists between two qubits only when their information statuses are 1. undetermined; 2. related by a number of QIEs that specify the conditions their values must satisfy when determined in some given bases. We have already seen the entangled example of $|\varphi_1\rangle$ and will discuss more entangled examples below that satisfy this information-logic criterion. States that do not satisfy this criterion are not entangled. For example, if one or both qubits' information status is already determined in some bases, then the two qubits must be in a product state form $|\phi\rangle_1 \otimes |\varphi\rangle_2$ that is unentangled. In addition, in the state $\frac{1}{\sqrt{2}}(|00\rangle_{13} + |11\rangle_{13})(|00\rangle_{24} + |11\rangle_{24})$, although both $q_1$ and $q_2$'s information statuses are undetermined, they are not related by any QIE, and thus are also unentangled to each other (although they are entangled to other qubits).

**2.3 Understanding the dormant entanglement phenomenon with the QIL theory.** Next we use the QIL theory to study more complex entangled states having multiple qubits to better illustrate its usefulness. The well-known Greenberger–Horne–Zeilinger (GHZ) state $|GHZ\rangle = \frac{1}{\sqrt{2}}(|000\rangle_{123} + |111\rangle_{123})$ and the state $|\psi^{(3)}\rangle = \frac{1}{2}\left[(|00\rangle_{12} + |11\rangle_{12})|0\rangle_3 + (|01\rangle_{12} + |10\rangle_{12})|1\rangle_3\right]$ both have the property of "dormant entanglement" such that the entanglement of a 2-qubit subsystem can be activated or destroyed by the basis choice of measurement in an external system [48]. For example, if $q_3$ in $|GHZ\rangle$ is measured to be $|0\rangle$ in the current basis, the state becomes $|000\rangle_{123}$; if $q_3$ is measured to be $|1\rangle$ in the current basis, the state becomes $|111\rangle_{123}$. So in either case $q_1$ and $q_2$ are unentangled. However, if we Hadamard-rotate $q_3$ first, and then measure $|q_3\rangle = |0\rangle$, the state becomes $|\varphi_1\rangle = \frac{1}{\sqrt{2}}(|00\rangle_{12} + |11\rangle_{12})$; if we Hadamard-rotate $q_3$ first, and then measure $|q_3\rangle = |1\rangle$, the state becomes $|\varphi_2\rangle = \frac{1}{\sqrt{2}}(|00\rangle_{12} - |11\rangle_{12})$. So in either case $q_1$ and $q_2$ are entangled. From the perspective of the $q_1$-$q_2$ subsystem, its entanglement can be activated or destroyed by the basis choice of measurement in the external system of $q_3$. The situation of the $|\psi^{(3)}\rangle$ state is the opposite: $q_1$ and $q_2$ are entangled if $q_3$ is measured in the current basis, but if $q_3$ is measured in the Hadamard-rotated basis, the possibility of entanglement is permanently destroyed.

To understand these properties with the QIL theory, we first construct the QIEs: The GHZ state can be created by applying the CNOT gate $CX_{2\to3}$ to the state $|\varphi_1\rangle_{12} \otimes |0\rangle_3$ ($CX_{2\to3}$ means $q_2$ controls $q_3$), so its QIE in the current basis is:

$$(q_1(c) \oplus q_2(c) = 0, \text{ and } q_2(c) \oplus q_3(c) = 0) \tag{5}$$

where applying $CX_{2\to3}$ effectively creates a duplicate of $q_2(c)$ on $q_3(c)$. If we measure $q_3$ in the current basis, we determine $q_3(c)$'s value, which by Eq. (5) implies $q_1(c)$ and $q_2(c)$ are both deterministic, i.e. $q_1$ and $q_2$ must have deterministic values when measured in the current basis, so $q_1$ and $q_2$ cannot be entangled anymore. However, if we measure $q_3$ in the Hadamard-rotated basis, we determine $q_3(h)$'s value and thus lose $q_3(c)$'s value, so the 2$^{nd}$ part of Eq. (5), $q_2(c) \oplus q_3(c) = 0$, no longer affects the value of $q_2(c)$, and only the 1$^{st}$ part $q_1(c) \oplus q_2(c) = 0$ is effective, so we reduce to the QIE in Eq. (2), which describes a Bell state – indeed measuring $|q_3\rangle = |0\rangle$ in the Hadamard-rotated basis transforms the $q_1$ - $q_2$ subsystem into $|\varphi_1\rangle = \frac{1}{\sqrt{2}}(|00\rangle_{12} + |11\rangle_{12})$, while measuring $|q_3\rangle = |1\rangle$ transforms $q_1$ - $q_2$ into $|\varphi_2\rangle = \frac{1}{\sqrt{2}}(|00\rangle_{12} - |11\rangle_{12})$, so the entanglement of $q_1$-$q_2$ is activated.

In the other example, the $|\psi^{(3)}\rangle$ state can be created by applying the gate sequence $CX_{3\to 2}H_3$ to the state $|\varphi_1\rangle_{12} \otimes |0\rangle_3$, so its QIE in the current basis is:

$$q_1(c) \oplus q_2(c) \oplus q_3(c) = 0 \tag{6}$$

where applying $H_3$ and then $CX_{3\to 2}$ effectively adds $q_3(c)$ to the existing $q_1(c) \oplus q_2(c)$. If we measure $q_3$ in the current basis, we determine $q_3(c)$'s value and transform Eq. (6) into $q_1(c) \oplus q_2(c) = q_3(c) = $ deterministic value, which describes a Bell state – indeed measuring $|q_3\rangle = |0\rangle$ in the current basis transforms the $q_1$ - $q_2$ subsystem into $|\varphi_1\rangle = \frac{1}{\sqrt{2}}(|00\rangle_{12} + |11\rangle_{12})$, while measuring $|q_3\rangle = |1\rangle$ transforms $q_1$-$q_2$ into $|\varphi_3\rangle = \frac{1}{\sqrt{2}}(|01\rangle_{12} + |10\rangle_{12})$, so the entanglement of $q_1$-$q_2$ is activated. However, if we measure $q_3$ in the Hadamard-rotated basis, we determine $q_3(h)$'s value and thus lose $q_3(c)$'s value. Because in Eq. (6) the values of $q_1(c)$ and $q_2(c)$ depend on $q_3(c)$, and once $q_3(c)$ is lost, both $q_1(c)$ and $q_2(c)$ are lost too. Now by complementarity, $q_1(h)$ and $q_2(h)$ are both deterministic, i.e. $q_1$ and $q_2$ must have deterministic values when measured in the Hadamard-rotated basis, and thus $q_1$ and $q_2$ cannot be entangled anymore.

The complete QIEs for $|GHZ\rangle$ and $|\psi^{(3)}\rangle$ in both the "$c$" and "$h$" bases are:

$$|GHZ\rangle\text{'s QIE:} \quad \begin{cases} (q_1(c) \oplus q_2(c) = 0,\text{ and } q_2(c) \oplus q_3(c) = 0) \\ q_1(h) \oplus q_2(h) \oplus q_3(h) = 0 \end{cases}$$

$$|\psi^{(3)}\rangle\text{'s QIE:} \quad \begin{cases} q_1(c) \oplus q_2(c) \oplus q_3(c) = 0 \\ (q_1(h) \oplus q_2(h) = 0,\text{ and } q_2(h) \oplus q_3(h) = 0) \end{cases} \tag{7}$$

In Eq. (7) the QIEs of $|GHZ\rangle$ and $|\psi^{(3)}\rangle$ are exactly the same if "c" and "h" are switched, which implies $|GHZ\rangle$ and $|\psi^{(3)}\rangle$ can be transformed into each other by applying Hadamard gates on all the qubits – consequently they are Pontryagin duals in the wave-particle duality of qubit states [26, 27]. Here using the QIEs we have easily identified the duality between $|GHZ\rangle$ and $|\psi^{(3)}\rangle$ without going through the algebra of applying the Hadamard gates and collecting terms. In general, if a state is created by only Hadamard, X, and CNOT gates, we can determine its QIE with relative ease: CNOT gates create duplicates or add new qubits into the equation, Hadamard changes between "c" and "h", X negates the value of the equation. Below we call any state that can be created by these gates a "Bell class entanglement". Clearly, all the four Bell states, $|GHZ\rangle$ and dormant entanglement states like $|\psi^{(3)}\rangle$ are entanglements of the Bell class. Despite the simple collection of gates allowed for its creation, a Bell class entanglement can have complex properties and applications as discussed below.

**2.4 The dormant entanglement paradox and its resolution by the QIL theory.** The dormant entanglement phenomenon leads to an apparent paradox which we will discuss below and use the QIL theory to resolve. Firstly, we create $|\psi^{(3)}\rangle$ and then move $q_3$ to far away location that it takes light considerable time to travel to the $q_1$-$q_2$ subsystem. Then we take another qubit $q_4 = |0\rangle$ at $q_3$'s location and entangle it to $q_3$ by the gate sequence $CX_{4\to 3}H_4$:

$$|\psi^{(4)}\rangle = CX_{4\to 3}H_4\left(|\psi^{(3)}\rangle_{123} \otimes |0\rangle_4\right) = \frac{1}{2\sqrt{2}}\begin{pmatrix}\left[(|00\rangle_{12} + |11\rangle_{12})|0\rangle_3 + (|01\rangle_{12} + |10\rangle_{12})|1\rangle_3\right]|0\rangle_4 \\ + \left[(|00\rangle_{12} + |11\rangle_{12})|1\rangle_3 + (|01\rangle_{12} + |10\rangle_{12})|0\rangle_3\right]|1\rangle_4\end{pmatrix} \tag{8}$$

By working out the algebra of $|\psi^{(4)}\rangle$, we find that $q_1$-$q_2$ can be transformed into a Bell state by measuring both $q_3$ and $q_4$ in the current basis: measuring $|q_3q_4\rangle = |00\rangle_{34}$ and $|q_3q_4\rangle = |11\rangle_{34}$ gives $|q_1q_2\rangle = \frac{1}{\sqrt{2}}(|00\rangle_{12} + |11\rangle_{12})$ ; while measuring $|q_3q_4\rangle = |01\rangle_{34}$ and $|q_3q_4\rangle = |10\rangle_{34}$ gives $|q_1q_2\rangle = \frac{1}{\sqrt{2}}(|01\rangle_{12} + |10\rangle_{12})$. On the other hand, if we measure $q_4$ in the Hadamard-rotated basis (details in the Supplementary Information (SI) Section S1), then no matter what happens to $q_3$,

8the possibility of getting an entanglement for $q_1$-$q_2$ is permanently destroyed: e.g. applying $H_4$ to $|\psi^{(4)}\rangle$ and measuring $|q_4\rangle=|0\rangle$ gives (after some algebra) $|q_1 q_2\rangle=|++\rangle_{12}$; measuring $|q_4\rangle=|1\rangle$ gives $|q_1 q_2\rangle=|--\rangle_{12}$; therefore both cases are product states (not entangled). So measuring $q_4$ in different bases has an instantaneous effect on the $q_1$-$q_2$ subsystem, which is similar to the spooky action of the Bell states. However, a critical difference is, in the Bell state spooky action, $q_1$'s entangling with $q_2$ happened locally, and only $q_1$'s measurement happened remotely from $q_2$. Here not only $q_4$'s measurement but also its entangling with $q_3$ happened remotely from the $q_1$-$q_2$ subsystem, so it is puzzling how $q_4$'s measurement can instantaneously determine if $q_1$ and $q_2$ are entangled or not.

This apparent paradox can be explained by the QIL theory. In Eq. (6) we have the QIE of $|\psi^{(3)}\rangle$ in the current basis. Now the way we added $q_4$ to $|\psi^{(3)}\rangle$ by $CX_{4\to 3}H_4$ is the same as when we added $q_3$ to $|\varphi_1\rangle$ by $CX_{3\to 2}H_3$, so the QIE of $|\psi^{(4)}\rangle$ in the current basis is just:

$$q_1(c)\oplus q_2(c)\oplus q_3(c)\oplus q_4(c)=0 \qquad (9)$$

Then by a logic similar to the one described in Section 2.3, if we measure $q_4$ in the current basis and determine $q_4(c)$, Eq. (9) becomes $q_1(c)\oplus q_2(c)\oplus q_3(c)=q_4(c)=$"deterministic value", which has the same structure of Eq. (6), and thus the entanglement of $q_1$ and $q_2$ can be activated by further measuring $q_3$ in the current basis. On the other hand, if we measure $q_4$ in the Hadamard-rotated basis, we determine $q_4(h)$'s value and thus lose $q_4(c)$'s value. Because in Eq. (9) the values of $q_1(c)$ and $q_2(c)$ have become dependent on $q_4(c)$, once $q_4(c)$ is lost, both $q_1(c)$ and $q_2(c)$ are lost too. Now by complementarity, $q_1(h)$ and $q_2(h)$ are both deterministic such that $q_1$ and $q_2$ must have deterministic values when measured in the Hadamard-rotated basis, and thus these two qubits cannot be entangled anymore. We see that although $q_4$'s interaction with $q_3$ happened remotely from the $q_1$-$q_2$ subsystem, it adds $q_4$ to the overall QIE involving $q_1$ and $q_2$, and thus measuring $q_4$ changes the information status of $q_1$ and $q_2$: this change happens instantaneously because in the QIL theory, the information status is a global property not restricted by the speed of light. Now by the discussion in the end of Section 2.2, whether $q_1$ and $q_2$ are entangled or not is decided by their information statuses, therefore measuring $q_4$ has an instantaneous effect on the entanglement of the $q_1$-$q_2$ subsystem despite the fact $q_4$ was added remotely. Furthermore, by the QIL theory, the information availability is a local property restricted by the speed of light, so the information about whether $q_4$ has been measured, and if measured then has what value in which basis, has to travel within the speed of light to $q_1$-$q_2$ for any physical change to be observable at this subsystem – this obeys the no-communication theorem. Consequently, the QIL theory has resolved the apparent paradox by reasoning with the information status and the QIE.

**2.5 Designing exotic entangled states with the QIL theory.** So far we have seen the usefulness of the QIL theory for interpreting properties of entangled states. However, the QIL theory can also be used to understand and design exotic entangled states by manipulating the QIEs. In a recent study [48] we have proposed a state in which two qubits are entangled but have no correlation when measured in any arbitrary basis. Now using the QIL theory, we explain how this kind of exotic states can be designed systematically. We start with $|\psi^{(3)}\rangle_{123} \otimes |0\rangle_L$ and apply $CX_{2 \to L}$ to create a new state:

$$|\psi^{(3+L)}\rangle = CX_{2 \to L}|\psi^{(3)}\rangle_{123} \otimes |0\rangle_L = \frac{1}{2}\left[\left(|00\rangle_{12}|0\rangle_L + |11\rangle_{12}|1\rangle_L\right)|0\rangle_3 + \left(|01\rangle_{12}|1\rangle_L + |10\rangle_{12}|0\rangle_L\right)|1\rangle_3\right]$$
(10)

It can be verified algebraically that $q_1$ and $q_2$ in $|\psi^{(3+L)}\rangle$ are entangled, but have no correlation when measured in any arbitrary basis (see Ref. [48] or the SI Section S2). However, this algebraic process is quite complex and tedious. To explain this phenomenon by the QIL theory we start with the simpler case of $|\psi^{(3)}\rangle$ with its QIE for the current basis as defined in Eq. (6), and notice that $q_1$ and $q_2$ in $|\psi^{(3)}\rangle$ are not correlated in the current basis, but correlated in the Hadamard-rotated basis (see the SI Section S3 for the algebraic verification). By the QIL theory, if we first measure $q_1$ in the current basis, we determine $q_1(c)$'s value, then Eq. (6) becomes $q_2(c) \oplus q_3(c) = q_1(c) =$ "deterministic value", so without also measuring $q_3(c)$ we still cannot decide $q_2(c)$'s value. Similarly if we first measure $q_2$ in the current basis, we cannot decide $q_1(c)$'s value. Consequently $q_1$ and $q_2$ are not correlated in the current basis. However, if we first measure $q_1$ in the Hadamard-rotated basis, we decide $q_1(h)$'s value and lose $q_1(c)$'s value. Now because Eq. (6) is the only equation that specifies the values of $q_1(c)$ through $q_3(c)$, losing $q_1(c)$'s value means $q_2(c)$ and $q_3(c)$ are both lost. Then by complementarity, losing $q_2(c)$ means $q_2(h)$ is deterministic. Therefore measuring $q_1(h)$ also determines $q_2(h)$'s value, which means $q_1$ and $q_2$ are perfectly correlated in the Hadamard-rotated basis – this result also applies when $q_2(h)$ is measured first. Certainly we can also verify the above results algebraically, but the QIL offers a simpler and more intuitive interpretation that can be used for the design of exotic states like $|\psi^{(3+L)}\rangle$, which we will explain next.

Compared to $|\psi^{(3)}\rangle$, applying $CX_{2\to L}$ to $|\psi^{(3)}\rangle_{123} \otimes |0\rangle_L$ effectively creates a duplicate of $q_2(c)$'s information on $q_L(c)$, so the QIE of $|\psi^{(3+L)}\rangle$ in the current basis is:

$$(q_1(c) \oplus q_2(c) \oplus q_3(c) = 0 \text{ and } q_2(c) \oplus q_L(c) = 0) \tag{11}$$

The first part of Eq. (11) is the same as Eq. (6), but the extra part of $q_2(c) \oplus q_L(c) = 0$ here can make a difference. We first consider the correlation of $q_1$ and $q_2$ in the current basis. Similar to the previous discussion on $|\psi^{(3)}\rangle$, measuring either $q_1(c)$ or $q_2(c)$ first cannot decide the other one's value, and the extra part $q_2(c) \oplus q_L(c) = 0$ does not provide any additional information on $q_1(c)$ or $q_2(c)$. However, when we consider the correlation of $q_1$ and $q_2$ in the Hadamard-rotated basis, then $q_2(c) \oplus q_L(c) = 0$ makes an important difference as compared to Eq. (6). In particular, if we now measure $q_1(h)$ and lose $q_1(c)$'s value, $q_2(c)$'s value is not lost this time, because an extra copy of $q_2(c)$'s information has been created on $q_L(c)$! That is, although the loss of $q_1(c)$ means we can no longer obtain $q_2(c)$ from the first part $q_1(c) \oplus q_2(c) \oplus q_3(c) = 0$, the second part $q_2(c) \oplus q_L(c) = 0$ still keeps open the possibility of determining $q_2(c)$ from $q_L(c)$. In the QIL language, Nature knows that $q_2(c)$'s information still exists and could still be obtained by its relation to $q_L(c)$, such that it will not determine the information status of $q_2$ in $q_2(h)$. Consequently, $q_1$ and $q_2$ are not correlated in the Hadamard-rotated basis.

Now what if we rotate $q_1$ or $q_2$ into an arbitrary basis? Suppose we rotate and measure $q_1$ first. The idea is that measuring $q_1$ in the Hadamard-rotated basis determines $q_1(h)$ and completely loses $q_1(c)$'s information, while measuring $q_1$ in any other basis causes a partial loss of information in $q_1(c)$, and the amount of information loss is greater when the unitary transformation is closer to the Hadamard gate. By $q_1(c) \oplus q_2(c) \oplus q_3(c) = 0$, any partial loss of information in $q_1(c)$ causes a partial loss of information in $q_2(c)$. However, due to the relation of $q_2(c) \oplus q_L(c) = 0$, $q_2(c)$'s information still exists in whole from $q_L(c)$. Because the information status is a global property, $q_2(c)$'s information existing in whole by its relation with $q_L(c)$ means its status is unchanged by any information change of $q_1$ from a global perspective, and consequently measuring $q_1$ first in any arbitrary basis has no effect on the measurement probabilities of $q_2$. Now suppose we measure $q_2$ first in an arbitrary basis, will the partial loss of $q_2(c)$ cause any change of the information status of $q_1$? No, because $q_2(c) \oplus q_L(c) = 0$ actually means $q_2(c) = q_L(c)$ so Eq. (11) actually contains a hidden equation of $q_1(c) \oplus q_L(c) \oplus q_3(c) = 0$. This means that $q_1(c)$'s information still exists in whole and can be obtained if both $q_L(c)$ and $q_3(c)$ are measured in the current basis. So measuring $q_2$ first in any arbitrary basis has no effect on the measurement probabilities of $q_1$. Consequently, due to the existence of $q_L(c)$ as a duplicate of $q_2(c)$, measuring either one of $q_1$ and $q_2$ in whatever basis has no effect on the information status of the other qubit, so indeed $q_1$ and $q_2$ are completely uncorrelated when measured in any arbitrary pair of bases. This is quite a remarkable result considering that $q_1$ and $q_2$ are indeed entangled (see Ref. [48] or the SI Section S2). Compared to the algebraic verification detailed in the SI, the QIL interpretation of $|\psi^{(3+L)}\rangle$ is much simpler and more intuitive. In addition, the above discussion furthers demonstrates the fundamental value of the global information status as introduced in Section 2.2.

In the above, knowing the exotic state $|\psi^{(3+L)}\rangle$, the QIL theory has been used to provide a simple and intuitive interpretation for it. However, without knowing $|\psi^{(3+L)}\rangle$ in the first place, we can also use the QIL theory to design this exotic state. Suppose our goal is to design a special quantum state in which two qubits $q_1$ and $q_2$ are entangled, but not correlated in any basis. For $q_1$ and $q_2$ to be entangled, they must be related by some QIE like Eq. (2) or Eq. (6). For $q_1$ and $q_2$ to be uncorrelated in the current basis, the QIE must not allow the determination of $q_1(c)$ to also determine $q_2(c)$, or vice versa, so $q_1(c) \oplus q_2(c) \oplus q_3(c) = 0$ is a good starting point. However, $q_1(c) \oplus q_2(c) \oplus q_3(c) = 0$ is not enough because measuring $q_1(h)$ causes the loss of $q_1(c)$, which then leads to the loss of $q_2(c)$ and the determination of $q_2(h)$. So the question becomes how to create a state for which the loss of $q_1(c)$ does not cause any loss of $q_2(c)$: the answer is we need to use another qubit to hold a duplicate of $q_2(c)$ to prevent its information loss when $q_1$ is measured in any basis. This logic naturally leads to the QIE of Eq. (11), which can be realized by applying $CX_{2 \to L}$ to $|\psi^{(3)}\rangle_{123} \otimes |0\rangle_L$. We see that by the QIL theory, the design of the exotic quantum state $|\psi^{(3+L)}\rangle$ becomes a logical process.

**2.6 Describing multi-qubit correlation behaviors with the QIL theory.** Next we discuss how the QIL theory really shines when used to describe the correlation behaviors of all qubits in a multi-qubit entanglement. The QIEs such as Eqs. (9) and (11) completely describe how all four qubits should correlate when measured in the current basis and the Hadamard-rotated basis.

Consider $|\psi^{(4)}\rangle$ as described by Eq. (9), obviously all four qubits are equivalent in this QIE and thus any pair of qubits are correlated in the same manner as all other possible pairs, so we only need to consider the pair of $q_1$-$q_2$. Firstly, measuring in the current basis, determining $q_1(c)$ first does not determine $q_2(c)$, and vice versa. In fact, $q_1(c)$ and $q_2(c)$ can only become correlated when both $q_3$ and $q_4$ are measured in the current basis, i.e. when $q_3(c)$ and $q_4(c)$ are determined such that the Eq. (9) reduces to $q_1(c)+q_2(c)=$"deterministic value" which corresponds to a Bell state. By the equivalence of all qubits, any pair of qubits in $|\psi^{(4)}\rangle$ are not correlated in the current basis until all other qubits are measured in the current basis. Secondly, if any qubit is measured in the Hadamard-rotated basis, its $q(h)$ is determined while $q(c)$ is lost – then by Eq. (9) all other qubits have lost their $q(c)$ values while determined their $q(h)$ values. In other words, measuring any one qubit in the Hadamard-rotated basis determines the values of all other qubits in the same basis, so any pair of qubits are perfectly correlated in the Hadamard-rotated basis.

Next consider $|\psi^{(3+L)}\rangle$ as described by Eq. (11), clearly $q_1$ and $q_3$ are equivalent, while $q_2$ and $q_L$ are equivalent. So we only need to consider the correlation behavior of three different pairs: $q_1$ and $q_3$, $q_2$ and $q_L$, $q_1$ and $q_2$, because any other possible pair is equivalent to one of these three. Firstly, by Eq. (11), measuring in the current basis, determining $q_1(c)$ first does not determine $q_2(c)$, and vice versa, so $q_1$ and $q_2$ are not correlated – however, they do become correlated when $q_3$ is measured in the current basis. Similarly, determining $q_1(c)$ first does not determine $q_3(c)$, and vice versa, so $q_1$ and $q_3$ are not correlated – however, they do become correlated when either $q_2$ or $q_L$ is measured in the current basis. Determining $q_2(c)$ automatically determines $q_L(c)$, and vice versa, so $q_2$ and $q_L$ are correlated in the current basis. Secondly, if $q_1$ is measured in the Hadamard-rotated basis, then $q_1(c)$ is lost while $q_1(h)$ is determined, then $q_3(c)$ is lost while $q_3(h)$ is determined, so $q_1$ and $q_3$ are correlated in the Hadamard-rotated basis. However, if $q_2(c)$ is lost by measuring $q_2$ in the Hadamard-rotated basis, $q_L(c)$ is not lost, because it can still be determined from its relation to $q_1$ and $q_3$ in Eq. (11), therefore $q_2$ and $q_L$ are not correlated in the Hadamard-rotated basis. Finally for $q_1$ and $q_2$, we have seen in Section 2.5 and SI Section S2 that they are not correlated in any arbitrary basis, including the Hadamard-rotated basis.

By the above arguments we see that the QIEs such as Eqs. (9) and (11) together with the QIL reasoning provide descriptions of quite complex correlation behaviors of all qubits in multi-qubit entanglements such as $|\psi^{(4)}\rangle$ or $|\psi^{(3+L)}\rangle$. Indeed, the correlation between qubits is directly determined by how one qubit's information status change affects other qubits' information status, and is not directly determined by whether or how much they are entangled – this is consistent with the result in Ref. [25] that entanglement is not directly related to the measurement statistics of qubits. Consequently, the information-logic-based QIL theory has a natural advantage in describing multi-qubit correlations over conventional entropy-based entanglement theories.

## 3. Conclusion

In this work we have introduced the qubit information logic (QIL) theory that uses the qubit information equation (QIE) and logic to describe correlations in multi-qubit entanglements. Starting with the Bell states, we introduced the global information status and local information availability that provided an intuitive interpretation of the spooky action and the quantum no-communication theorem. We then proceeded to use the QIL and QIE to study the dormant entanglement phenomenon as illustrated by the GHZ state and $\left|\psi^{(3)}\right\rangle$. We next discussed an apparent paradox caused by the dormant entanglement and resolved it with the QIL and QIE. Furthermore, we demonstrated the usefulness of the QIL and QIE by using them to understand and design an exotic entanglement state $\left|\psi^{(3+L)}\right\rangle$ in which two qubits are entangled but not correlated in any arbitrary basis. Finally, we showed the QIL and QIE have a natural advantage over conventional theories when used to describe the correlation behaviors of all qubits in a multi-qubit entanglement: because the correlations between qubits are directly determined by how one qubit's information status change affects other qubits' information status, and are not directly determined by whether or how much they are entangled. Overall, the new theory of QIL and QIE provides a new understanding of entanglement based on information logic and correlation of measurements, and can be used to interpret and design exotic quantum states.

## 4. Acknowledgements

The authors acknowledge the funding from the U.S. Department of Energy (Office of Basic Energy Sciences) under Award No. DE-SC0019215, and from the NSF grant 2124511 [CCI Phase I: NSF Center for Quantum Dynamics on Modular Quantum Devices (CQD-MQD)].

## 5. Supplementary Information is available after the References.

**Supplementary Information: The qubit information logic theory for understanding multi-qubit entanglement and designing exotic entangled states**

Zixuan Hu and Sabre Kais*

*Department of Chemistry, Department of Physics, and Purdue Quantum Science and Engineering Institute,*
*Purdue University, West Lafayette, IN 47907, United States*
*\*Email:* kais@purdue.edu

## S1. The algebraic derivation for the dormant entanglement state $|\psi^{(4)}\rangle$ as discussed in the main text Section 2.4

In the main text Section 2.4 the dormant entanglement paradox is illustrated with the state $|\psi^{(4)}\rangle$:

$$|\psi^{(4)}\rangle = CX_{4\to 3}H_4\left(|\psi^{(3)}\rangle_{123}\otimes|0\rangle_4\right) = \frac{1}{2\sqrt{2}}\begin{pmatrix}\left[\left(|00\rangle_{12}+|11\rangle_{12}\right)|0\rangle_3 + \left(|01\rangle_{12}+|10\rangle_{12}\right)|1\rangle_3\right]|0\rangle_4 \\ +\left[\left(|00\rangle_{12}+|11\rangle_{12}\right)|1\rangle_3 + \left(|01\rangle_{12}+|10\rangle_{12}\right)|0\rangle_3\right]|1\rangle_4\end{pmatrix} \quad S(1)$$

In the main text we have seen the $q_1$-$q_2$ subsystem can be transformed into a Bell state by measuring both $q_3$ and $q_4$ in the current basis. Here we will present the algebraic derivation that shows if we measure $q_4$ in the Hadamard-rotated basis, then no matter what happens to $q_3$, the possibility of getting an entanglement for $q_1$-$q_2$ is permanently destroyed. We first rotate $q_4$ by the Hadamard gate and then collect the terms:

$$\begin{aligned}H_4|\psi^{(4)}\rangle &= \frac{1}{2\sqrt{2}}\begin{pmatrix}\left[\left(|00\rangle_{12}+|11\rangle_{12}\right)|0\rangle_3 + \left(|01\rangle_{12}+|10\rangle_{12}\right)|1\rangle_3\right](|0\rangle+|1\rangle)_4 \\ +\left[\left(|00\rangle_{12}+|11\rangle_{12}\right)|1\rangle_3 + \left(|01\rangle_{12}+|10\rangle_{12}\right)|0\rangle_3\right](|0\rangle-|1\rangle)_4\end{pmatrix} \\ &= \begin{pmatrix}\begin{bmatrix}\left(|00\rangle_{12}+|11\rangle_{12}+|01\rangle_{12}+|10\rangle_{12}\right)|0\rangle_3 \\ \left(|01\rangle_{12}+|10\rangle_{12}+|00\rangle_{12}+|11\rangle_{12}\right)|1\rangle_3\end{bmatrix}|0\rangle_4 \\ +\begin{bmatrix}\left(|00\rangle_{12}+|11\rangle_{12}-|01\rangle_{12}-|10\rangle_{12}\right)|0\rangle_3 \\ \left(|01\rangle_{12}+|10\rangle_{12}-|00\rangle_{12}-|11\rangle_{12}\right)|1\rangle_3\end{bmatrix}|1\rangle_4\end{pmatrix} \quad S(2) \\ &= \left(|++\rangle_{12}|0\rangle_3+|++\rangle_{12}|1\rangle_3\right)|0\rangle_4 + \left(|--\rangle_{12}|0\rangle_3-|--\rangle_{12}|1\rangle_3\right)|1\rangle_4 \\ &= \left(|++\rangle_{12}|+\rangle_3\right)|0\rangle_4 + \left(|--\rangle_{12}|-\rangle_3\right)|1\rangle_4\end{aligned}$$

17By Eq. S(2), if we now measure $q_4$, then if $|q_4\rangle=|0\rangle$ we have: $(|++\rangle_{12}|+\rangle_3)|0\rangle_4$, if $|q_4\rangle=|1\rangle$ we have: $(|--\rangle_{12}|-\rangle_3)|1\rangle_4$. So in either case, the $q_1$-$q_2$ subsystem has become a product state that is no longer entangled, regardless if $q_3$ is measured or not. Consequently measuring $q_4$ in the Hadamard-rotated basis instantaneously destroy the entanglement of $q_1$-$q_2$, which leads to the paradox described in Section 2.4 in the main text.

**S2. The algebraic derivation of how $q_1$ and $q_2$ in $|\psi^{(3+L)}\rangle$ are entangled but have no correlation when measured in any arbitrary basis, as discussed in the main text Section 2.5**

In the main text Section 2.5 around Equation (10) we have stated that $q_1$ and $q_2$ in $|\psi^{(3+L)}\rangle$ are entangled but have no correlation when measured in any arbitrary basis. The algebraic verification of this property can be found in our recent work (arXiv:2306.05517, (2023)) which is also summarized below.

$|\psi^{(3+L)}\rangle$ is defined in the main text to be:

$$|\psi^{(3+L)}\rangle = \frac{1}{2}\left[(|00\rangle_{12}|0\rangle_L + |11\rangle_{12}|1\rangle_L)|0\rangle_3 + (|01\rangle_{12}|1\rangle_L + |10\rangle_{12}|0\rangle_L)|1\rangle_3\right] \qquad S(3)$$

Firstly $q_1$ and $q_2$ in $|\psi^{(3+L)}\rangle$ are entangled because applying a Hadamard gate to $q_L$ in $|\psi^{(3+L)}\rangle$ gives:

$$H_L|\psi^{(3+L)}\rangle = \frac{1}{2\sqrt{2}}\begin{bmatrix}(|00\rangle_{12}(|0\rangle_L+|1\rangle_L)+|11\rangle_{12}(|0\rangle_L-|1\rangle_L))|0\rangle_3 \\ +(|01\rangle_{12}(|0\rangle_L-|1\rangle_L)+|10\rangle_{12}(|0\rangle_L+|1\rangle_L))|1\rangle_3\end{bmatrix}$$

$$= \frac{1}{2\sqrt{2}}\begin{bmatrix}((|00\rangle_{12}+|11\rangle_{12})|0\rangle_L+(|00\rangle_{12}-|11\rangle_{12})|1\rangle_L)|0\rangle_3 \\ +((|01\rangle_{12}+|10\rangle_{12})|0\rangle_L-(|01\rangle_{12}-|10\rangle_{12})|1\rangle_L)|1\rangle_3\end{bmatrix} \qquad S(4)$$

$H_L$ is the Hadamard gate applied on $q_L$

By Equation S(4), measuring $q_3$ and $q_L$ now make $q_1$ and $q_2$ into a Bell state: $|0\rangle_L|0\rangle_3$ gives $\frac{1}{\sqrt{2}}(|00\rangle_{12}+|11\rangle_{12})$, $|1\rangle_L|0\rangle_3$ gives $\frac{1}{\sqrt{2}}(|00\rangle_{12}-|11\rangle_{12})$, $|0\rangle_L|1\rangle_3$ gives $\frac{1}{\sqrt{2}}(|01\rangle_{12}+|10\rangle_{12})$, $|1\rangle_L|1\rangle_3$ gives $\frac{1}{\sqrt{2}}(|01\rangle_{12}-|10\rangle_{12})$. So if $q_1$ and $q_2$ were not already entangled in $|\psi^{(3+L)}\rangle$, then their entanglement would have been created by local operations on $q_3$ and $q_L$, which is impossible. Consequently $q_1$ and $q_2$ are already entangled in $|\psi^{(3+L)}\rangle$.

Secondly to verify $q_1$ and $q_2$ are not correlated when measured in any basis, we first rotate the two qubits by arbitrary unitary transformations:

$$U_1 U_2 \left|\psi^{(3+L)}\right\rangle = \frac{1}{2}\begin{pmatrix}\left[\left(a_1|0\rangle_1 + a_2|1\rangle_1\right)\left(b_1|0\rangle_2 + b_2|1\rangle_2\right)|0\rangle_L + \left(a_2^*|0\rangle_1 - a_1^*|1\rangle_1\right)\left(b_2^*|0\rangle_2 - b_1^*|1\rangle_2\right)e^{i(\alpha+\beta)}|1\rangle_L\right]|0\rangle_3 \\ + \left[\left(a_1|0\rangle_1 + a_2|1\rangle_1\right)\left(b_2^*|0\rangle_2 - b_1^*|1\rangle_2\right)e^{i\beta}|1\rangle_L + \left(a_2^*|0\rangle_1 - a_1^*|1\rangle_1\right)\left(b_1|0\rangle_2 + b_2|1\rangle_2\right)e^{i\alpha}|0\rangle_L\right]|1\rangle_3\end{pmatrix}$$

$$U_1 = \begin{pmatrix} a_1 & a_2^* e^{i\alpha} \\ a_2 & -a_1^* e^{i\alpha} \end{pmatrix} \otimes I \otimes I \otimes I, \quad U_2 = I \otimes \begin{pmatrix} b_1 & b_2^* e^{i\beta} \\ b_2 & -b_1^* e^{i\beta} \end{pmatrix} \otimes I \otimes I, \quad |a_1|^2 + |a_2|^2 = |b_1|^2 + |b_2|^2 = 1$$

S(5)

In Equation S(5) $U_1$ and $U_2$ are arbitrary basis transformations applied to $q_1$ and $q_2$ respectively. If we do not measure $q_1$ first, the probability of measuring $|q_2\rangle = |0\rangle$ in $U_1 U_2 \left|\psi^{(3+L)}\right\rangle$ is:

$$p(|q_2\rangle = |0\rangle) = \frac{1}{4}\left[\left(|a_1|^2 + |a_2|^2\right)|b_1|^2 + \left(|a_2^*|^2 + |a_1^*|^2\right)|b_2^*|^2 + \left(|a_1|^2 + |a_2|^2\right)|b_2^*|^2 + \left(|a_2^*|^2 + |a_1^*|^2\right)|b_1|^2\right]$$
$$= \frac{1}{2}$$

S(6)

If we measure $q_1$ first, suppose we measure $|q_1\rangle = |0\rangle$, then $U_1 U_2 \left|\psi^{(3+L)}\right\rangle$ will collapse into the state:

$$U_1 U_2 \left|\psi^{(3+L)}\right\rangle (|q_1\rangle = |0\rangle) = \frac{1}{\sqrt{2}}\begin{pmatrix}\left[a_1|0\rangle_1\left(b_1|0\rangle_2 + b_2|1\rangle_2\right)|0\rangle_L + a_2^*|0\rangle_1\left(b_2^*|0\rangle_2 - b_1^*|1\rangle_2\right)e^{i(\alpha+\beta)}|1\rangle_L\right]|0\rangle_3 \\ + \left[a_1|0\rangle_1\left(b_2^*|0\rangle_2 - b_1^*|1\rangle_2\right)e^{i\beta}|1\rangle_L + a_2^*|0\rangle_1\left(b_1|0\rangle_2 + b_2|1\rangle_2\right)e^{i\alpha}|0\rangle_L\right]|1\rangle_3\end{pmatrix}$$

S(7)

And then the conditional probability of measuring $|q_2\rangle = |0\rangle$ given $|q_1\rangle = |0\rangle$ is:

$$p(|q_2\rangle = |0\rangle \mid |q_1\rangle = |0\rangle) = \frac{1}{2}\left(|a_1 b_1|^2 + |a_2^* b_2^*|^2 + |a_1 b_2^*|^2 + |a_2^* b_1|^2\right) = \frac{1}{2} = p(|q_2\rangle = |0\rangle) \quad \text{S(8)}$$

Consequently by Equation S(8) we have $p(|q_2\rangle = |0\rangle \mid |q_1\rangle = |0\rangle) = p(|q_2\rangle = |0\rangle)$ which means $q_1$ and $q_2$ have independent values when measured in arbitrary basis. So we have verified that $q_1$ and $q_2$ are not correlated when measured in any arbitrary basis. We see that this algebraic verification is quite complex and tedious as compared to the QIL arguments presented in the main text Section 2.5.

**S3. The algebraic derivation of how $q_1$ and $q_2$ in $\left|\psi^{(3)}\right\rangle$ are not correlated when measured in the current basis, but correlated in the Hadamard-rotated basis, as discussed in the main text Section 2.5.**

$\left|\psi^{(3)}\right\rangle$ is defined in the main text to be:

$$\left|\psi^{(3)}\right\rangle = \frac{1}{2}\left[\left(|00\rangle_{12} + |11\rangle_{12}\right)|0\rangle_3 + \left(|01\rangle_{12} + |10\rangle_{12}\right)|1\rangle_3\right] \qquad \text{S(9)}$$

Firstly to verify $q_1$ and $q_2$ are not correlated when measured in the current basis, if $q_1$ is not measured first, the probability of measuring $q_2 = |0\rangle$ is $p(q_2 = |0\rangle) = \frac{1}{4} + \frac{1}{4} = \frac{1}{2}$. If $q_1$ is measured first, suppose we measure $|q_1\rangle = |0\rangle$, then $\left|\psi^{(3)}\right\rangle$ will collapse into the state:

$$\left|\psi^{(3)}\right\rangle(|q_1\rangle = |0\rangle) = \frac{1}{2}\left(|00\rangle_{12}|0\rangle_3 + |01\rangle_{12}|1\rangle_3\right)$$
$$= \frac{1}{2}|0\rangle_1\left(|0\rangle_2|0\rangle_3 + |1\rangle_2|1\rangle_3\right) \qquad \text{S(10)}$$

By Eq. S(10) we have conditional probability of measuring $|q_2\rangle = |0\rangle$ given $|q_1\rangle = |0\rangle$ is:

$$p(|q_2\rangle = |0\rangle | |q_1\rangle = |0\rangle) = \frac{1}{4} + \frac{1}{4} = \frac{1}{2} = p(|q_2\rangle = |0\rangle) \qquad \text{S(11)}$$

Consequently we have $p(|q_2\rangle = |0\rangle | |q_1\rangle = |0\rangle) = p(|q_2\rangle = |0\rangle)$ which means $q_1$ and $q_2$ are independent and thus not correlated in the current basis.

Secondly to verify $q_1$ and $q_2$ are correlated when measured in the Hadamard-rotated basis, we apply the Hadamard gate to both $q_1$ and $q_2$, we get:

$$H_1 H_2 \left|\psi^{(3)}\right\rangle = \frac{1}{4}\begin{pmatrix}\left[(|0\rangle_1 + |1\rangle_1)(|0\rangle_2 + |1\rangle_2) + (|0\rangle_1 - |1\rangle_1)(|0\rangle_2 - |1\rangle_2)\right]|0\rangle_3 \\ + \left[(|0\rangle_1 + |1\rangle_1)(|0\rangle_2 - |1\rangle_2) + (|0\rangle_1 - |1\rangle_1)(|0\rangle_2 + |1\rangle_2)\right]|1\rangle_3\end{pmatrix}$$
$$= \frac{1}{2}\left((|00\rangle_{12} + |11\rangle_{12})|0\rangle_3 + (|00\rangle_{12} - |11\rangle_{12})|1\rangle_3\right) \qquad \text{S(12)}$$

$$H_1 = H \otimes I \otimes I \qquad H_2 = I \otimes H \otimes I$$

By Equation S(12), $q_1$ and $q_2$ are perfectly correlated when measured in the $\{|+\rangle, |-\rangle\}$ basis because they must be both $|0\rangle$ or both $|1\rangle$.